\documentstyle[preprint,aps,epsf]{revtex}

\begin{document}
\draft
\title{Scalar field cosmology in three-dimensions.}
\author{G. Oliveira-Neto\thanks{Email: gilneto@fisica.ufjf.br}}
\address{Departamento de Fisica,
Instituto de Ciencias Exatas,
Universidade Federal de Juiz de Fora,
CEP 36036-330, Juiz de Fora,
Minas Gerais, Brazil.}
\date{\today}
\maketitle
\begin{abstract}
We study an analytical solution to the Einstein's equations 
in $2+1$-dimensions. The space-time is dynamical and has a
line symmetry. The matter content is a minimally 
coupled, massless, scalar field. Depending on the value of 
certain parameters, this solution represents three distinct 
space-times. The first one is flat space-time.
Then, we have a big bang model with a negative 
curvature scalar and a real scalar field. The last case is a 
big bang model with event horizons where the curvature scalar
vanishes and the scalar field changes from real to purely 
imaginary.
 
\end{abstract}

\section{Introduction.}
\label{sec:introduction}

Many recent works have focused attention on the issue of general
relativity in $2+1$-dimensions \cite{jackiw}. Most of them deal
with two important issues: black holes and cosmology.

The main motivation for the most recent works in black hole 
physics came from the important discovery of the BTZ black hole
\cite{banados}. They investigate different classical and quantum 
properties of the BTZ and other black hole solutions
\cite{mann}, \cite{carlip}.

On the other hand, the simplicity of the theory in 
three-dimensions, when compared with its version in 
four-dimensions, is the main motivation for the works in
cosmology. Most of them deal with quantum aspects of
the theory, in particular the attempt to derive a wave-function
for the Universe  \cite{carlip}, \cite{gil1}, \cite{louko}.

In the present work we would like to study an analytical solution
to the Einstein's equations in $2+1$-dimensions. The space-time
is dynamical and has a line symmetry. The matter content
is represented by a minimally coupled, massless, scalar field.
We shall see that this solution gives rise, depending
on the value of certain parameters, to three different
space-times. The first one is flat space-time.
Then, we have a big bang model with a negative 
curvature scalar and a real scalar field. The last case is a 
big bang model with event horizons where the curvature scalar
vanishes and the scalar field changes from real to purely imaginary.

In Sec. \ref{sec:equations}, we start by writing down the appropriate
Einstein's equations. A solution ($S$) of this set of equations has 
already been found, in different circumstances Ref. \cite{wang}. We
shall use $S$ suitably adapted to our problem.

$S$ is composed of few functions of the relevant coordinates and free
parameters ($p_i$). In Sec. \ref{sec:solutions}, we impose conditions
upon $S$ such that it may be interpreted as a physically acceptable
solution. These conditions restrict the domains of the $p_i$'s, and
divide $S$ in three distinct types of space-times, depending on the
values of the $p_i$'s. The space-times are the ones mentioned above
and we study them in great details.

Finally, in Sec. \ref{sec:conclusions} we summarize the main points 
and results of the paper.

\section{Einstein-scalar equations.}
\label{sec:equations}

We shall start by writing down the ansatz for the $2+1$-dimensional,
dynamical, line symmetric, space-time metric as,

\begin{equation}
\label{1}
ds^2\, =\, -\, 2 e^{(N(u,v) + U(u,v)/2)} du dv\, +\, e^{-2U(u,v)}
dy^2\, .
\end{equation}
where $N(u,v)$ and $U(u,v)$ are two arbitrary functions to be
determined by the field equations, $(u,v)$ is a pair of null
coordinates varying in the range $(-\infty,\infty)$, and $y$
is a coordinate taking values over a real line:  
$(-\infty,\infty)$.

The scalar field $\varphi$ will be a function only of the two null
coordinates and the expression for its stress-energy tensor
$T_{\alpha\beta}$ is given by \cite{wheeler},

\begin{equation}
\label{2}
T_{\alpha\beta}\, =\, \varphi,_\alpha \varphi,_\beta\, -\,
{1\over 2} g_{\alpha\beta} \varphi,_\lambda \varphi^{,_\lambda}\, .
\end{equation}
where $,$ denotes partial differentiation.

Now, combining Eqs. (\ref{1}) and (\ref{2}) we may obtain the
Einstein's equations, which in the units of Ref. \cite{wheeler}
take the following form,

\begin{equation}
\label{3}
2 U,_{uu}\, -\, (U,_u)^2\, +\, 2 N,_u U,_u\, =\, 2 (\varphi,_u)^2\, ,
\end{equation}
\begin{equation}
\label{4}
2 U,_{vv}\, -\, (U,_v)^2\, +\, 2 N,_v U,_v\, =\, 2 (\varphi,_v)^2\, ,
\end{equation}
\begin{equation}
\label{5}
2 N,_{uv}\, +\, U,_{uv}\, =\, 2 \varphi,_u \varphi,_v\, ,
\end{equation}
\begin{equation}
\label{6}
(e^U),_{uv}\, =\, 0\, .
\end{equation}
The equation of motion for the scalar field, in these 
coordinates, is

\begin{equation}
\label{7}
2 \varphi,_{uv}\, -\, U,_u \varphi,_v\, -\, U,_v \varphi,_u\, =\, 
0\, ,
\end{equation}

Comparing the above set of non-linear, second-order, coupled,
partial, differential equations (\ref{3})-(\ref{7}) with the
one derived in Ref. \cite{wang}, called type $A$, we notice that they
are identical, although the situation treated here is different from
the one treated there. Therefore, we shall take as the solution of the
system above Eqs. (\ref{3})-(\ref{7}), the type $A$ space-times,  
obtained in Ref. \cite{wang}. They will be introduced and studied
in great details in the next section.

\section{Solutions.}
\label{sec:solutions}

From our discussion in the previous section, we suitably adapt the
type $A$ space-times of Ref. \cite{wang} to the present situation,
and write the following solution to the system Eqs. 
(\ref{3})-(\ref{7}),

\begin{equation}
\label{8}
\exp{(-U(u,v))}\, \equiv \, t^2(u,v)\, =\, (1-u^{2\eta}-v^{2\mu})^2\, ,
\end{equation}
\begin{eqnarray}
\label{9}
\exp{(-N(u,v))} \equiv W(u,v)\, =&& t^{2\delta} \left[ {u^\eta
v^\mu - (1-v^{2\mu})^{1/2} (1-u^{2\eta})^{1/2} \over u^\eta v^\mu
+ (1-v^{2\mu})^{1/2} (1-u^{2\eta})^{1/2} }\right]^{2\gamma}
\left[ {u^\mu - (1-v^{2\mu})^{1/2} \over u^\eta + (1 - 
v^{2\mu})^{1/2} }\right]^{4a\delta_+}\nonumber\\
&&\times 
\left[ {v^\mu - (1 - 
u^{2\eta})^{1/2} \over v^\mu + (1 - u^{2\eta})^{1/2} 
}\right]^{4a\delta_-}\, (1-v^{2\mu})^{-\delta^2_+}\, 
(1-u^{2\eta})^{-\delta^2_-}\, ,
\end{eqnarray}
\begin{eqnarray}
\label{10}
\varphi(u,v)\, =&& 2a\ln{t^2}\, +\, \rho_+ \ln{\left[ {1 - 
u^\eta (1-v^{2\mu})^{1/2} - v^\mu (1-u^{2\eta})^{1/2} \over
1 + u^\eta (1-v^{2\mu})^{1/2} + v^\mu (1-u^{2\eta})^{1/2}}
\right] }\nonumber\\
&&+
\rho_- \ln{\left[ {1 - 
u^\eta (1-v^{2\mu})^{1/2} + v^\mu (1-u^{2\eta})^{1/2} \over
1 + u^\eta (1-v^{2\mu})^{1/2} - v^\mu (1-u^{2\eta})^{1/2}}
\right] }\,
\end{eqnarray}
where $\mu$ and $\eta$ $\in$ $\Re$ and they are greater than or
equal to $1/2$, $\delta = 2a^2 + (3 - 1/\eta - 1/\mu )/4$,
$\gamma^2 = (1 - 1/2\eta )(1 - 1/2\mu )$, $\delta^2_+ = 1 - 1/2\eta$,
$\delta^2_- = 1 - 1/2\mu$, $\rho_\pm = (1/2) (\delta_+ \pm 
\delta_- )$, and $a$ is a constant, real, number.

In terms of the new quantities $t(u,v)$ Eq. (\ref{8}), and $W(u,v)$
Eq. (\ref{9}), the line element Eq. (\ref{1}) becomes,

\begin{equation}
\label{11}
ds^2\, =\, - 2 W(u,v) t^{1/2}(u,v) du dv\, +\, t^2(u,v) dy^2\, ,
\end{equation}
One may notice from Eqs. 
(\ref{8})-(\ref{10}), that for different values of $\eta$, $\mu$ 
and $a$, one has different space-times. It is also important to
notice that the choice of the letter $t$ in Eq. (\ref{8}), was
not casual. We shall be able to interpret it as a time coordinate.

Observing Eq. (\ref{11}), we notice that these space-times have a
singularity at $t=0$.
It is a physical singularity as can be seen directly from the
curvature scalar $R$, and also from Ref. \cite{wang}.

In order to show this result we start writing down the Ricci tensor
that, in the present case, has the following expression \cite{taub},

\begin{equation}
\label{12}
R_{\alpha \beta}\, =\, \varphi,_\alpha\, \varphi,_\beta \, .
\end{equation}
From it, we may compute $R$ straightforwardly with the aid of
Eqs. (\ref{8})-(\ref{11}), finding,

\begin{equation}
\label{13}
R\, =\, {-8 \mu \eta v^{2\mu -1} u^{2\eta -1} ( 4a + \delta_+ +
\delta_- )^2 \over W t^{5/2}}\, .
\end{equation}

Finally, taking the limit $t \to 0$ in $R$ Eq. (\ref{13}), we
find that this quantity diverges at $t=0$. There is no other physical
singularity for these space-times because $R$ is well defined outside
$t=0$. We can also learn that taking the limit $t \to \infty$ of $R$
Eq. (\ref{13}), this quantity goes to zero.

The space-times above will only be physically acceptable if $t$ 
Eq. (\ref{8}) is a real, positive function and $W$ Eq. (\ref{9}) 
is a real function. Therefore, only few distinct sets of values of
$\mu$, $\eta$ and $a$ will be allowed. Each of them giving
rise to a different type of space-time. It is important to
note that since $t$ and $W$ are functions of ($u$,$v$),
the range of these coordinates shall also be restricted.

Here we shall loose the condition that the scalar field
$\varphi$ Eq. (\ref{10}), be real. We shall permit it to be
purely imaginary in some space-time regions, which means
that in those regions $\varphi$ will be an example of the
so-called `exotic matter' \cite{thorne}.

\subsection{Space-times.}
\label{subsec:space-times}

We start now the determination of the allowed values of $\mu$,
$\eta$ and $a$ by imposing that $t$ be a real, positive,
function. In order to simplify our study we shall demand that
$2\mu$ and $2\eta$ be integers, greater than or equal to $1$.

Observing the definition of $t$ Eq. (\ref{8}), we notice that
if we permit $2\mu$ or $2\eta$ to be even, for positive $t$, we
would have very limited domains for the coordinates $u$ and $v$,
respectively. Since we would like to have the biggest possible
domains for these coordinates, we shall restrict our 
attention to odd, integer, values of $2\mu$ and $2\eta$. The
positiveness of $t$ will also select the physically accessible
space-time volume.

From the expression of W Eq. (\ref{9}), we learn that the
three distinct functions of $u$ and $v$, inside brackets,
respectively with exponents $2\gamma$, $4a\delta_+$ and 
$4a\delta_-$, may be negative even for positive $t$. Since,
in order to interpret $t$ as a time coordinate, $W$ has to be 
positive (see Eq. (\ref{13,5}) below), we shall eliminate these
terms. The best way to accomplish this is by setting theirs 
exponents to zero.

With the aid of the formulae for $\gamma$, $\delta_+$ and
$\delta_-$, given just below Eq. (\ref{10}), we understand
that there are three distinct manners to set the above
exponents to zero: (a) by choosing $a = 0$ and $2\mu = 2\eta
= 1$, (b) by choosing $2\mu = 2\eta = 1$, and finally (c)
by choosing $a = 0$ and either $2\mu$ or $2\eta$ equal to $1$.
As we shall see below, these possibilities will produce the
distinct sets of space-times associated to our solution.

Now, we are in the position to identify $t$ as a time 
coordinate. For positive $W$ and $t$, we may compute the
norm of the quadri-vector normal to surfaces of constant $t$.
Which gives,

\begin{equation}
\label{13,5}
- 8 \eta \mu u^{2\eta - 1} v^{2\mu -1} {1\over W t^{1/2}},
\end{equation}
which is always negative for $2\eta$ and $2\mu$, odd, integer,
numbers. 

Gathering together all the conditions obtained above, we may
group the solutions satisfying these conditions in three
distinct sets.

\subsubsection{Flat space-time.}
\label{subsubsec:flat}

The first is empty, flat space-time, obtained for
$2\mu = 2\eta = 1$ and $a = 0$. It has the following line
element, from Eqs. (\ref{8}), (\ref{9}) and (\ref{11}),

\begin{equation}
\label{14}
ds^2\, =\, - 2 du dv\, +\, t^2 dy^2\, ,
\end{equation}
where, $t = 1 - u - v$.

For positive $t$, the coordinates $u$ and $v$ will vary in
the range ($-\infty$ , $\infty$), but under the condition,

\begin{equation}
\label{15}
u\, +\, v\, <\, 1\, .
\end{equation}

The surface $t = 0$ in this case is not a physical
singularity, it is just a coordinate singularity.

\subsubsection{Big bang cosmology without horizons.}
\label{subsubsec:bbnohorizon}

The solutions belonging to this set have $2\mu = 2\eta = 1$ and
$a \neq 0$. Introducing these values in Eqs. (\ref{8}), 
(\ref{9}) and (\ref{11}), we may write the following line
element,

\begin{equation}
\label{18}
ds^2\, =\, - 2 t^{4 a^2} du dv\, +\, t^2 dy^2\, ,
\end{equation}
where $t = 1 - u - v$, $u$, $v$ $\in$ ($-\infty ,\infty$) and
satisfy Eq. (\ref{15}).

The scalar field Eq. (\ref{10}) is now,

\begin{equation}
\label{19}
\varphi\, =\, 2 a \ln{ t^2 }\, .
\end{equation}

In the present case, we may see from Eq. (\ref{19}) that the
scalar field is always real, therefore the stress-energy 
Eq. (\ref{2}) is always positive.

The scalar of curvature Eq. (\ref{13}), is written

\begin{equation}
\label{20}
R\, =\, - {32 a^2 \over t^{(4a^2 + 2)}}\, .
\end{equation}
From this expression is easy to see that $t = 0$ is still a
physical singularity for these space-times. The scalar field
Eq. (\ref{19}), is also singular there. Therefore, we may
interpret this singularity as a big bang, for this space-time.

The dynamical nature of this space-time may be better
appreciated if we re-write the line element Eq. (\ref{18})
in terms of $t$ and $x = u - v$,

\begin{equation}
\label{21}
ds^2\, =\, {t^{4a^2}\over 2} (\, -\, dt^2\, +\, dx^2\, )\,
+\, t^2 dy^2,
\end{equation}
where $-\infty < x < \infty$.

\subsubsection{Big bang cosmology with event horizons.}
\label{subsubsec:bbhorizon}

The last set of space-times is determined when we set 
$a = 0$ and either $2\mu$ or $2\eta$ equal to $1$. As a
matter of definition, and without loosing the generality,
let us choose $2\eta = 1$ and $2\mu = 2n + 1$, where $n$ is a
positive integer.

With the aid of Eqs. (\ref{8}), (\ref{9}) and (\ref{11}),
the line element of these space-times is,

\begin{equation}
\label{22}
ds^2\, =\, - 2 \left( {t\over 1 - u}\right)^{({2n\over 2n 
+ 1})} du dv\, +\, t^2 dy^2\, ,
\end{equation}
where $t = 1 - u - v^{2n + 1}$, $u$, $v$ $\in$ ($-\infty 
,\infty$) and satisfy the condition $u + v^{2n + 1} < 1$.

Observing Eq. (\ref{22}), we identify besides the singularity
at $t = 0$, another one at $u = 1$. The second singularity is
not a physical one as can be seen directly from $R$ Eq. 
(\ref{13}), which for the present situation is,

\begin{equation}
\label{23}
R\, =\, {- 4 n v^{2n} (1 - u)^{({2n\over 2n + 1})}\over
t^{({6n + 2\over 2n +1})}}\, .
\end{equation}
Indeed, $u = 1$ is an event horizon as we shall
demonstrate below.

Since $u = 1$ is just a coordinate singularity, there are
new coordinates which let Eq. (\ref{22}) regular at this
event. On the other hand, in order to better understand 
the physical effect of the horizon and the dynamical 
nature of the space-time, it is more convenient to 
re-write the line element Eq. (\ref{22}) in terms of,
$t$ and $x\, =\, -\, 1\, +\, u\, - v^{2n+1}$.
Which gives,

\begin{equation}
\label{27}
ds^2\, =\, {1\over 4n+2}\left({4t\over x^2
-t^2}\right)^{\left({2n\over 2n+1}\right)}(\, -\, 
dt^2\, +\, dx^2\, )\, +\, t^2 dy^2\, .
\end{equation}
Here, it is clear that we have not only the $u =0$
horizon, which in the new coordinates is the surface
$t =x$, but another one at $t = -x$. In the old 
coordinates this is the surface $v =0$. 

The basic property of event horizons, is the fact
that they isolate certain space-time regions from another
ones \cite{wheeler}. This can be demonstrated for
the surfaces $t = \pm x$, in the following way if we
restrict our attention to the $(t, x)$ plane.

Let us start by calling {\it sector I}, the 
region in the past of the horizons ($0 < t < \pm x$, 
$-\infty < x < \infty$). We also introduce the {\it sector
II}, as the region to the future of the horizons 
($\pm x < t < \infty$, $-\infty < x < \infty$).
From Eq. (\ref{27}) we can see that null rays describe
$45^\circ$ or $135^\circ$ straight lines in the ($t,
x$) plane. Therefore, not even the light will be able to 
return from {\it sector II} to {\it sector I}, 
once it has entered it. Although $t$ does
not change the role of time with $x$ after crossing the
horizons, as we shall see few important facts take 
place there.

The scalar field Eq. (\ref{10}), may be obtained for the
space-times being studied, in the new set of coordinates.

\begin{equation}
\label{28}
\varphi(t ,x)\, =\, {1\over 2} \sqrt{{2n\over 2n + 1}} 
\ln{\left[ {(x\, +\, ( x^2 - t^2 )^{1/2}
\over (x\, -\, ( x^2 - t^2 )^{1/2}}\right]}\, .
\end{equation}

If one inspects the expression of $\varphi(t,x)$ Eq. 
(\ref{28}), one notices that for $t=0$ it diverges.
Therefore, also for the present space-times we may
interpret $t=0$ as a big bang singularity.

In {\it sector I}, $\varphi(t,x)$ Eq. (\ref{28})
is a real function. On the other hand, in {\it sector II}
the scalar field is purely imaginary. It means that
in {\it sector II}, the stress-energy Eq. (\ref{2}) 
may be negative and the scalar field will be an example of
the so-called `exotic matter'.

Observing $R$ Eq. (\ref{23}) we see that it vanishes
in both horizons but has the same sign in {\it sectors I} and
{\it II}.

Finally, If one were interested in studying quantum field 
theory in {\it sector II}, the resulting
theory would be unitary. This is the case because there is
no singularities there and, from Eq. (\ref{13,5}), the 
space-times under investigation possess a global timelike 
Killing vector field \cite{davies}.

\section{Conclusions.}
\label{sec:conclusions}

In the present work we have studied an analytical solution
to the Einstein's equations in $2+1$-dimensions. The space-time
was dynamical and had a line symmetry. The matter content
was represented by a minimally coupled, massless, scalar field.

The Einstein's equations for this system were identical to 
another one already solved in the literature. We have used the
known solution ($S$), and studied it.

We have imposed certain conditions upon $S$ such that it
could be interpreted as a physically acceptable solution.
These conditions restricted the domains of few free parameters
($p_i$) of $S$, and divided $S$ in three distinct types of
space-times, depending on the values of the $p_i$'s.
The first one was flat space-time.
Then, we had a big bang model with a negative 
curvature scalar and a real scalar field. The last case was a 
big bang model with event horizons where the curvature scalar
vanishes and the scalar field changes from real to purely imaginary.

\acknowledgments

I am grateful to A. Wang for suggestive discussions in the
course of this work. I would like also to thank
I. D. Soares for helpful discussions and FAPEMIG for the
invaluable financial support.

\end{document}